\documentclass[journal]{IEEEtran}
\IEEEoverridecommandlockouts
\usepackage{times}
\usepackage{flushend}
\usepackage{cite}
\usepackage[dvips]{color}
\usepackage{epsf}
\usepackage{epsfig}
\usepackage{graphicx}
\usepackage{graphics}
\usepackage{epstopdf}
\usepackage[footnotesize]{caption}
\usepackage[bottom]{footmisc}
\usepackage{amsmath}
\usepackage{amssymb}
\usepackage{amsthm}
\usepackage{amsxtra}
\usepackage{algorithm}
\usepackage{algorithmic} 	
\usepackage{enumerate}
\usepackage{multirow}
\usepackage{enumerate}
\usepackage{amssymb}
\usepackage{lipsum}
\usepackage{setspace}
\usepackage{multirow}
\usepackage{tikz}
\usepackage{setspace}
\usepackage{subfigure}
\usepackage{float}
\usepackage{caption}

\usepackage{rotating} 
\usepackage{graphicx} 

\newcommand{\Rmnum}[1]{\expandafter\@slowromancap\romannumeral #1@}

\makeatother
\begin{document}
\title{System Design of Internet-of-Things for Residential Smart Grid}
\author{Sanjana~Kadaba~Viswanath,~Chau~Yuen,~\IEEEmembership{Senior Member,~IEEE,}~Wayes~Tushar,~\IEEEmembership{Member,~IEEE,}~Wen-Tai~Li,~\IEEEmembership{Student Member,~IEEE,}~Chao-Kai~Wen,~\IEEEmembership{Member,~IEEE,}~Kun~Hu,~Cheng~Chen,~and~Xiang~Liu,~\IEEEmembership{Member,~IEEE}
\thanks{S. K. Viswanath, C. Yuen, W. Tushar and W.-T. Li are with the Singapore University of Technology and Design (SUTD), Somapah Road, Singapore 487372 (Email: \{sanjana, wayes\_tushar, yuenchau, wentai\_li\}@sutd.edu.sg).}
\thanks{C.-Kai Wen is with the National Sun Yat-Sen University, Taiwan (Email:~chaokai.wen@mail.nsysu.edu.tw).}
\thanks{Kun Hu, Cheng Chen and Xiang Liu are with the School of Software and Microelectronics, Peking University, Beijing, China (Email: \{hk19900116,~chengchen8901\}@163.com, xliu@ss.pku.edu.cn).}
\thanks{This work is supported in part by the Singapore University of Technology and Design through the Energy Innovation Research Program Singapore under Grant NRF2012EWT-EIRP002-045, in part by the SUTD-MIT International Design Center, Singapore under Grant IDG31500106, and in part by the Grant NSFC 61550110244.}
}
\IEEEoverridecommandlockouts
\maketitle
\begin{abstract}
Internet-of-Things (IoTs) envisions to integrate, coordinate, communicate, and collaborate real-world objects in order to perform daily tasks in a more intelligent and efficient manner. To comprehend this vision, this paper studies the design of a large scale IoT system for smart grid application, which constitutes a large number of home users and has the requirement of fast response time. In particular, we focus on the messaging protocol of a universal IoT home gateway, where our cloud enabled system consists of a backend server, unified home gateway (UHG) at the end users, and user interface for mobile devices. We discuss the features of such IoT system to support a large scale deployment with a UHG and real-time residential smart grid applications. Based on the requirements, we design an IoT system using the XMPP protocol, and implemented in a testbed for energy management applications. To show the effectiveness of the designed testbed, we present some results using the proposed IoT architecture.
\end{abstract}
 \setcounter{page}{1}
\section{Introduction}\label{sec:introduction}
The smart grid, with its two-way information and power flow capacities through ubiquitous interconnections of equipment in power networks, enables internet-of-things (IoT) to control and coordinate smart devices and thus paves the way towards energy management of large-scale systems~\cite{Ruilong-TII:2015,Tushar-TIE:2014,Liu:2014}. However, the smart devices need to have a set of capabilities including communication and cooperation, addressability,  identification, sensing, actuation, interoperability, embedded information processing, and user interfacing to achieve such seamless deployment to form the IoT. Further, to achieve these capabilities, a number of protocols are required to define communication pattern and software features between the devices as well as there is a need to determine approaches that would support the Internet's end-to-end functionality in order to create a user friendly smart technology with IoT~\cite{Liu-TSG:2015}. This leads us into a thorough research on setting up a IoT testbed that is targeting residential consumers to provide real-time information to consumers and keeping the latency to control IoT  devices to a minimum.

In this respect, this paper focuses on the IoT elements, protocols, and the testbed setup for IoT environments along with the protocols and software designs that have been used to monitor and control consumers' energy usage patterns. We have deployed smart home technology in a real world scenario with each housing unit with 3 bed rooms and a living room that can accommodate 6-9 consumers. Each unit consists of sensors, actuators, smart meters, smart plugs, a Universal Home Gateway (UHG), and together establish a home area network (HAN). Each of these smart devices communicates with the UHG through a different communication protocol. The UHG, on the other hand, interacts with cloud server where most of the processing is done. The implemented system can control and manage energy based on published dynamic pricing information, and can act as an energy management system. These capabilities of the UHG enable automatic demand response such as avoiding appliances usage during peak hours based on price signals and peak-load shaving; third party engagement in performing different tasks including energy and water monitoring, security and fire control for users; participation of residents in real-time energy market and building smart homes. Further, we have developed an Android mobile application to provide remote services to the consumers. Due to profound interaction of smart devices, we mainly emphasise on system design, IoT protocols and software implementation, which are essential for such lucrative deployment.

\section{State-of-The Art}\label{sec:state-of-art}
For the last few years, there has been a significant research interest in exploring the potential of IoT system design and their application in different aspect of our day-to-day life. Examples of such studies can be found in \cite{IoTServey:2010} and \cite{IoTServeystandard:2013}. Here, \cite{IoTServey:2010} provides a very broad overview of IoT and show how different technologies and communication technologies are merged together to benefit the everyday life. This study also depicts how different scientific domains view the applications of IoT from their own perspectives including 
\begin{enumerate}
\item Transportation and logistics such as logistics [Karpischek et al. (NFC-2009)], mobile ticketing [Broll et al. (IEEE IC-2009)] and efficient supply chain [Ilic et al. (IEEE PC-2009)].
\item Healthcare domain [Vilamovska et al. (RAND Europe-2009)].
\item Smart environment domain including home and offices [Buckl et al. (WAINA-2009)] and commercial buildings [Spiess et al. (IEEE ICWS-2009)].
\item Personal and social domain [Welbourne et al. (IEEE IC-2009)].
\end{enumerate} 
Please note that all the above mentioned references can be found in \cite{IoTServey:2010}, which we skip here due to the constraints on the reference number by the magazine. 

With a view to meet the important criteria of power efficiency, reliability and Internet connectivity, the authors in \cite{IoTServeystandard:2013} discusses different IoT standards of the IEEE $802.15.4$ (Standard for Information Technology Std., September 2006.) and IETF working groups\footnote{Available online in: http://www.ietf.org/dyn/wg/charter/core-charter.html.}. The authors introduce and relate key requirements of the power-efficient IEEE $802.15.4-2006$ PHY layer, the power saving  and reliable IEEE $802.15.4e$ MAC layer, the IETF $6$LoWPAN adaptation layer enabling universal Internet connectivity, the IETF ROLL routing protocol enabling availability, and finally the IETF CoAP enabling seamless transport and support of Internet applications. Further discussion on different aspects of IoT can also be found in \cite{Ruilong-TII:2015}, \cite{RongYu:2015}. and in studies, e.g, [Chang et al. (SmartGridComm-2012)] on IEEE standards\footnote{Examples of smart metering standards are available online in https://standards.ieee.org/findstds/standard/1377-2012.html and http://standards.ieee.org/develop/msp/smartgrid.pdf} in metering communications.

Now, based on the above discussion, our proposed study differs from the existing studies in following ways.
\begin{itemize}
\item Although there are studies on different applications of IoT and the standards in the literature, there is no detail study on any particular aspect related to the smart grid. In this paper, we zoom into the residential energy management aspect of smart grid and explored the potentiality of IoT in managing energy for residential customers.
\item Most of the works emphasize on the theoretical part of IoT system, and little has been done so far on the implementation aspect. We have complemented the existing studies on IoT by developing a testbed at SUTD and demonstrating its real-life applications.
\end{itemize}
To this end, we have discuss the details the detail of a IoT testbed, specially the idea of a unified home gateway (UHG) that we are currently prototyping. Based on the technology used, the proposed system set up also leverage to handle a large number of users whereby also possess a fast response time. 

\section{Elements of IoT System}\label{sec:ElementIoT}
IoT needs to abide to a number of elements, as explained below, to form a complete system. 
\begin{figure*}[t]
\centering
\includegraphics[width=\textwidth]{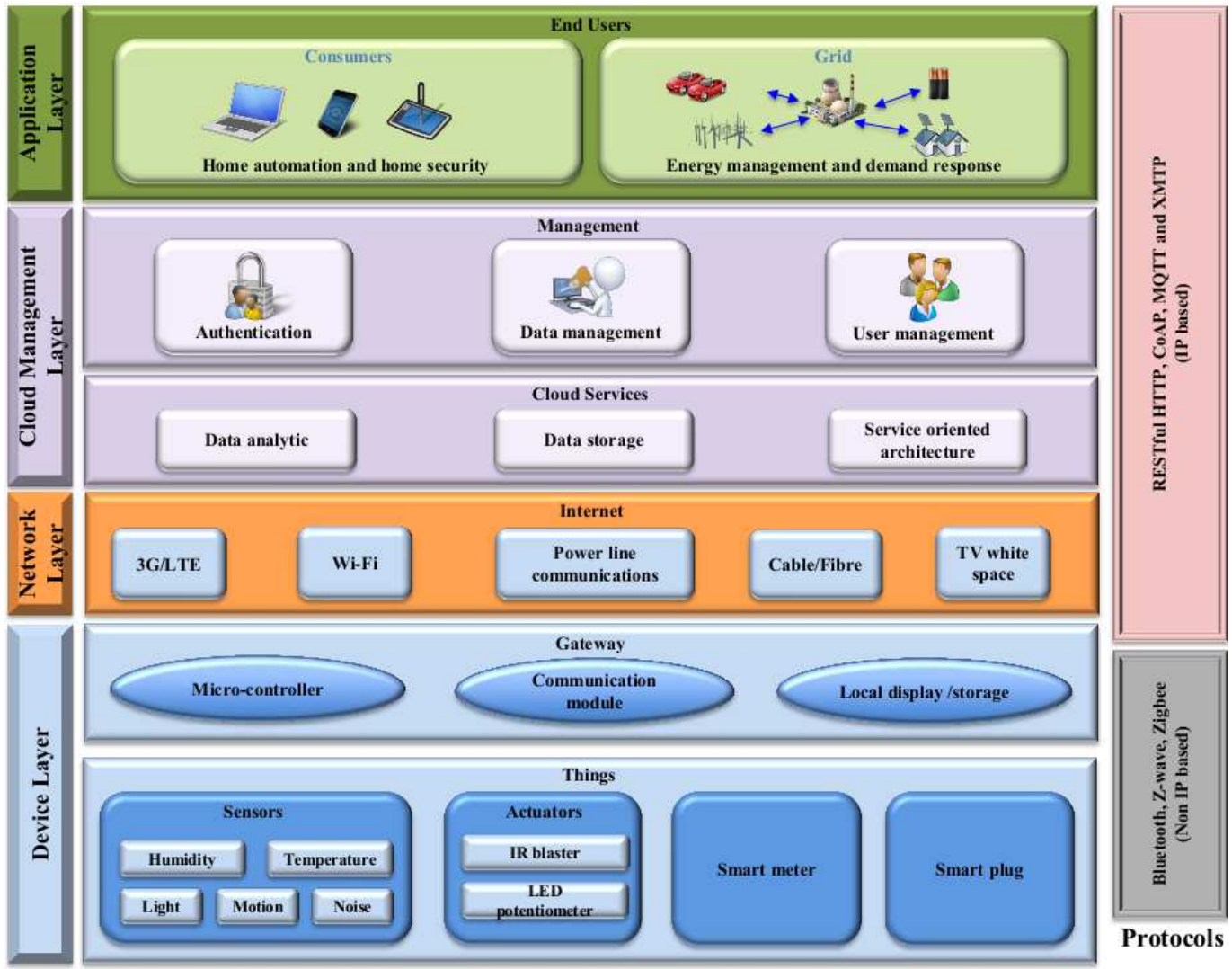}
\caption{IoT Layers.} \label{fig:iot}
\end{figure*}

\subsection{IoT Layers}
IoT can be divided into four major layers including device layer, network layer, cloud management layer, and application layer as shown in Fig.~\ref{fig:iot}.

\subsubsection{Device Layer}
This comprises of two sub-layers. Things layer consists of sensors, actuators, smart plugs and smart meters and is responsible for sensing environment, collecting data, and controlling home appliances. Gateway Layer hosts the micro-controller, communication module, local storage, and display, where components from things layer are connected to.

\subsubsection{Network Layer}
This layer connects device layer to application layer.

\subsubsection{Cloud Management Layer}
Cloud services layer is where all the data storage and information retrieval will take place. Management Layer is where authentication, user management and data management are done.

\subsubsection{Application Layer}This layer is responsible for providing services to the end users (home owner or smart grid). It hosts DRM, dynamic pricing for smart grid system, or Energy Management, Home Security for consumers services. The services can also be provided by third party, e.g. a home security management company.

\subsection{Comparison of IoT Protocols}
Since IoT comprises a large number of devices and communication intensive architecture, there is a need of standardize software protocols to enable all devices to communicate with one another with various features.

\subsubsection{RESTful HTTP}
The Representational State Transfer (REST) is an architectural style and not a protocol or standard.  RESTful works on top of HTTP protocol and uses the REST principles at the same time. Hence, RESTful HTTP is lightweight and has a simple HTTP request format and is very easy to implement. REST is best suited for applications where periodic communication is required. However, to increase the end user privacy, HTTPS can be used.

\subsubsection{CoAP}
Constrained Application Protocol (CoAP) is a specialized web transfer protocol for using with constrained devices and networks (e.g., low power, lossy). CoAP provides a request/response interaction model between application end points, supports  built-in  discovery  of  services  and  resources and  includes  key concepts of the web. CoAP is designed to easily interface  with  HTTP  for  integration  with  the  web  while  meeting  specialised requirements  such  as  multicast  support,  very  low  overhead  and  simplicity.

\subsubsection{MQTT}
Message Queue Telemetry Transport (MQTT) is publish/subscribe, extremely simple, and lightweight messaging protocol, designed for constrained devices and low bandwidth, high-latency or unreliable networks. The design principles are to minimise network bandwidth and device resource requirements whilst also attempting to ensure reliability and some degree of assurance of delivery. These principles also   make  the  protocol  ideal  for  the  emerging IoT world of connected devices with an objective of collecting data from multiple devices and transporting to IT infrastructure.

\subsubsection{XMPP}
The Extensible Messaging and Presence Protocol   (XMPP) is an application profile of the Extensible Markup Language (XML) that enables near-real-time exchange of structured yet extensible data between any two or more network entities~\cite{TusharWCM:2016}. XMPP is highly extensible through multiple  XMPP Extension Protocols (XEPs). XEPs provide various unique capabilities to XMPP such as interoperability, provisioning, security, scalability and low latency, which makes interaction with smart devices feasible and seamless. XMPP handles both simple and complex encoding of IoT data realising interoperability to a great extent and provides an interoperable way to publish control parameters and perform control operations. XMPP also enables publishing to multiple things under a single XMPP address through integration with subsystems.

\subsection{IoT Features for Universal Home Gateway}\label{sec:section4}
In this section, we will describe the features required for the universal home gateway in IoT system.
\subsubsection{Security}
Pervasive nature of IoT systems demands security to be a vital feature. As the use of smart devices is growing at an exponential pace, the need for strong security is becoming more important. For example, in 2013, a security researcher showed the security vulnerability of Philips IoT gadgets by hacking their secured Hue lighting system. As a result of these security attacks, consumers might receive wrong pricing information on their energy usage or a malicious controller, other than the authorised grid, may take the control over their scheduled loads. Therefore, strong security is required for connected devices to ensure data confidentiality, data integrity, and user/device authenticity. Hence, a large number of research efforts is on the privacy~\cite{Xu2015} and security issue in terms of both application layer data perturbation and secure communication~\cite{Keoh2014}.
\subsubsection{Provisioning}
Due to a great number of interconnected devices in IoT, making rules for devices communication and their access to services, data, and resources via provisioning is very important for management of energy. The success of a number of important aspects such as demand management with dynamic pricing, Home Automation system, and more importantly the security \cite{Keoh2014} significantly relies on proper provisioning of the smart devices within IoT.
\subsubsection{Interoperability}
Interoperability is a vital concept of IoT, specially in a smart home implementation, which let the devices in a network to connect through a common platform in order to work together. Smart grid is composed of large number of smart homes, where each home has a number of intelligent devices, that can operate, communicate and interact autonomously. Users may purchase an off-the-shelf devices, e.g., a sensor like camera or an actuator like light controller, and integrate them to the system. Interoperability of IoT is critical in making these smart devices work together.
\subsubsection{Latency}
IoT protocols need to have minimum latency for information retrieval and executing control messages as most IoT services related to energy management rely on real-time services and control information. For instance, in the event of an unexpected outage of scheduled generating plant, control signal to turn off the flexible loads needs to arrive within a duration of 8 seconds, 30 seconds or 10 minutes depending on whether primary, secondary or tertiary reserve power is used respectively\footnote{http://www.e-control.at/en/marketplayers/electricity/electricitymarket/
balancin\%20energy.}. However, it is really challenging to achieve a near real-time communication. Nevertheless, this can be achieved by a fully dedicated bidirectional, asynchronous communication channel with a capability to enable native push to the client.
\subsubsection{Massive number of devices}
Particularly in DRM, the utility can send information of its dynamic pricing or load control command, e.g., air conditioning (AC) system's thermal set point, to a large number of consumers to encourage them to avoid their consumption during peak hour or force their load to operate at certain power rating  during emergency. Similarly, an energy user with distributed renewable source can also communicate with the grid in order to determine the amount of energy it wants to buy from/sell to the grid at a particular time of the day.  Further, the ability to broadcast messages should be considered as it can deliver the messages to all the devices in one single message and hence not jam the network. Thus, the massive feature along with broadcast messaging, is essential for successful energy management application.

\subsubsection{Scalability} Different features of IoTs need to be highly scalable in order to provide services to large scale systems, i.e., the softwares and protocols should have designed such that new devices can be easily added to the system later, and yet fulfilling the strict quality-of-services such as delay requirements. This would leverage the extension of the smart grid network beyond a specific geographic area and number of smart devices, and also avoid the constraint on the number of smart devices that can be installed within a smart grid system for desired outcomes.
\begin{figure*}[t]
\centering
\includegraphics[width=\textwidth]{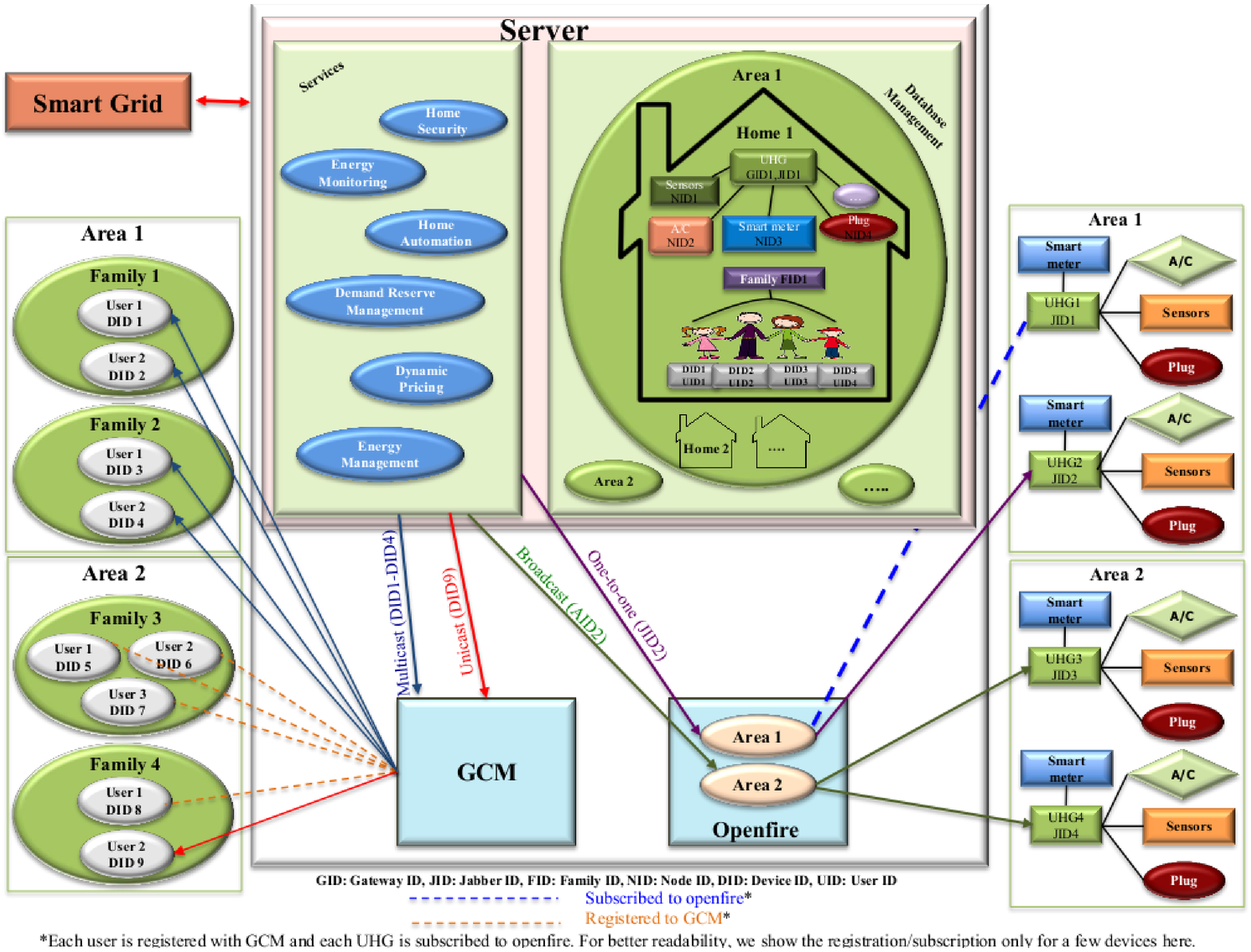}
\caption{Overview of the architecture, which intends to have multiple services in the cloud to provide different applications such as home automation services to the users, dynamic pricing module managed by the grid to perform demand shaping. Further, we can also have a third party to provide home security by analyzing the information collected through the system. All the protocol in the figure are for XMPP. The openfire is the open source XMPP used in our testbed, while the GCM is the XMPP server by Google for Android messaging. The data uploading of gateway to the cloud through REST API is not shown to avoid confusion.}
\label{fig:arch}
\end{figure*}

\section{Testbed Design for Residential Energy Management}\label{sec:section5}
In this section, we will describe the design of testbed focusing on residential household.
\subsection{Overall System Setup}
\subsubsection{Architecture}
The overall architecture of testbed targeting energy management applications is shown in Fig.~\ref{fig:arch}. Smart grid/smart home setup comprises of a cloud server where we store data as well as provide services for both gateways and consumers. Application layer and cloud management layer are implemented within the cloud server to simplify data sharing between the two layers. A number of services is developed to cater energy management and smart grid applications. Grid service handles all the commands from our smart grid and sends notifications either about DRM or dynamic pricing to consumers and gateways. Energy management service is responsible for monitoring and giving energy related information to consumers. Home Automation service is activated when a control command from consumers is received.

We have implemented push capability on top of Openfire server on the device side and Google Cloud Messaging (GCM) on the mobile apps side.  The incapability of Openfire to send real-time notifications to mobile app while offline drove us to implement a dual XMPP server with GCM and Openfire working together. 

We have grouped a number of homes (i.e., gateways and families) into an area according to their geographical location with an Area ID (AID). Each family comprises of a number of consumers is assigned a Family ID (FID). Each of their smart phones is registered with a unique Device ID (DID) issued by GCM and is used to push notifications to individual user. Each smart phone is assigned a unique User ID (UID) at the cloud server, which can be the user phone number that GCM is not aware of. Along with DID, UID can be used for authentication purpose. Thus, only authorised consumers can control their smart devices and monitor their own energy usage. Each UHG is assigned a unique Gateway ID (GID) by cloud server and a unique Jabber ID (JID) by Openfire. Each device in home network is assigned a Node ID (NID) through which the command is sent from UHG to end devices. GID is our cloud server's reference to the gateway and Openfire is not aware of GID. JID is Openfire reference to the gateway and cloud server uses it to push control commands during home automation where control is pushed only to a particular gateway (e.g. a user can only switch on/off a plug of his own house). Furthermore, during broadcasting of messages for demand response management (DRM), cloud server publishes a message to a particular area using AID to Openfire, which then broadcasts the message to all the UHGs corresponding to that AID.

\subsubsection{Network Protocols (XMPP and RESTful HTTP)}
In the testbed, RESTful HTTP is selected for periodic uploading of sensor data from the gateway to the cloud. Although there are many desirable properties of an IoT system, the proposed system mainly focuses on to improve the response time and to be able to handle a large number of users. Further, there are also needs to push control commands towards the end users for different applications such as DRM, which require asynchronous communication model. Please note that XMPP is not only capable of sending asynchronous request but can also support a massive number of users at the same time. Further, XMPP has fast response time, and provides security, provisioning and interoperability features required by energy management application \cite{Guo2015}. Furthermore, both iOS and Android mobile operating system have provided build in support for XMPP, which facilitate the connection to end user through the mobile apps. In this context, XMPP is selected as the network protocol to be used in the proposed system.

Please note that there could be a number of other protocol choices as explained in Section~\ref{sec:ElementIoT}, and selecting the right protocol is never an easy task. The reason behind not choosing CoAP in this work is that CoAP is UDP based, and typically UDP traffic is not as firewall friendly as TCP. Besides, due to the lack of TCPs retransmission mechanisms, packet loss is more likely to happen when using CoAP. Hence, we decided to use pub/sub based TCP protocol, which is important for the broadcast / multicast needs in smart grid application. While MQTT and XMPP are very similar to each another, we simply choose XMPP over MQTT as most mobile operating systems (e.g. Android or iOS) support XMPP for messaging. So our thought is to have a single messaging platform, specially since it is important to push message to user mobile device for the home gateway / smart grid application. Nonetheless, we claim no performance superiority of XMPP over CoAP and MQTT.

\subsection{Gateway Side}
\subsubsection{Openfire for massive number of devices, low latency and provisioning}
Openfire uses pubsub mechanism to push information to subscribers without providing configuration settings for pubsub services. To overcome this, we have designed and implemented our own configuration for provisioning devices and defining permissions for each of the devices e.g., distinguish publishers from subscribers. This was achieved by building a component that is an XMPP client in our cloud. The design goals of our component are threefold. Firstly, our component must be able to provision devices and assign different roles and responsibilities to each of them specifying their accessibilities.  Secondly, the component should listen to any incoming messages. Finally, it needs to send messages to pubsub service for publishing to all its subscribers (e.g., by broadcasting the message to all UHGs in an area) or to a single UHG (e.g., one-to-one communication for home automation). Our setup, where all UHG of same area subscribe to single area, enables grid to broadcast DRM or dynamic pricing messages to a massive number of residential household with a short delay.

\subsubsection{UHG for interoperability and scalability}
\begin{figure*}[t]
\centering
\includegraphics[width=\textwidth]{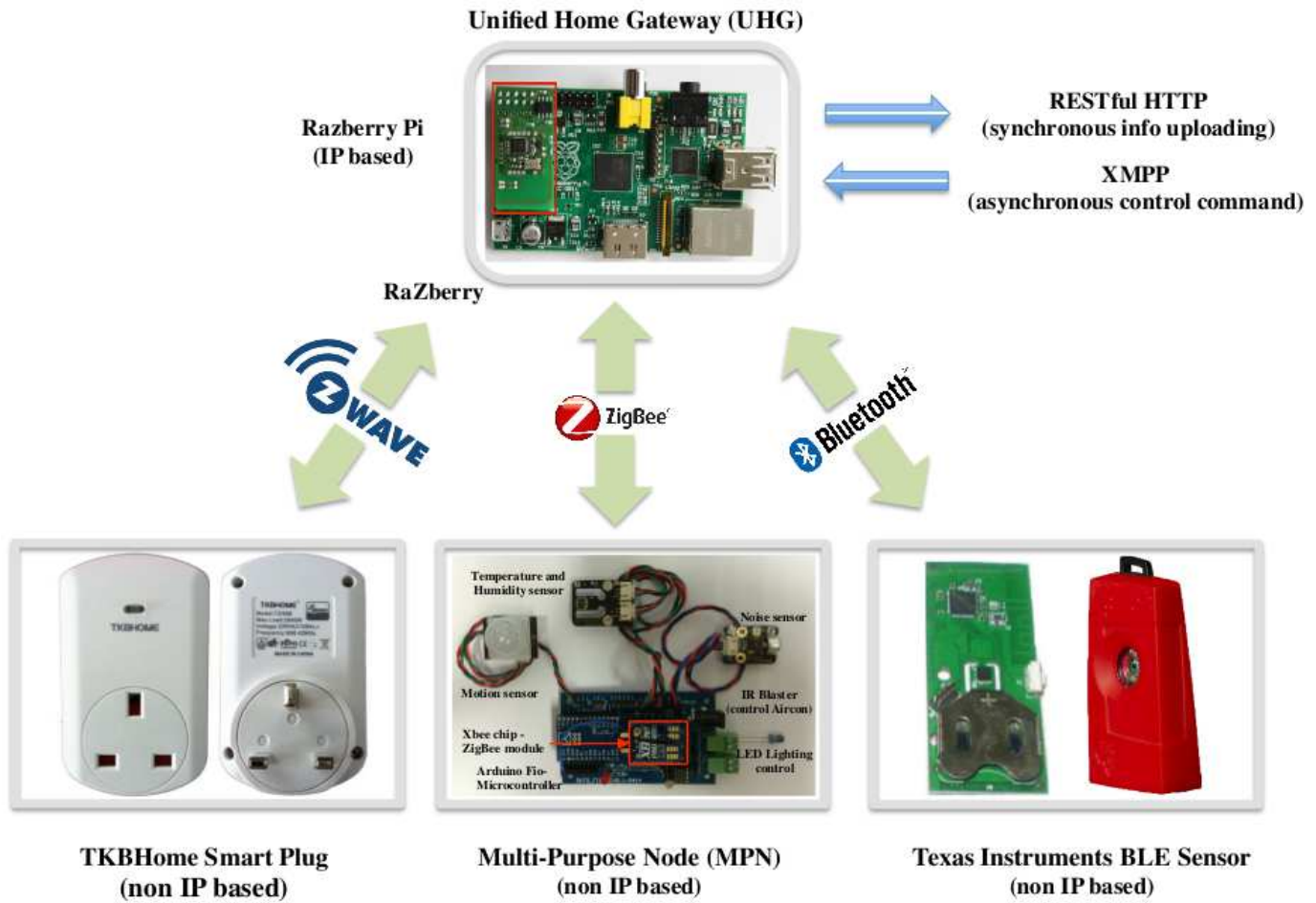}
\caption{Communication of the UHG with different non-IP based systems.}
\label{fig:msg}
\end{figure*}

UHG plays an important role in our implementation of smart home where interoperability and scalability is key. Each smart home unit accommodates a UHG, which communicates to all devices and sensors, and has a room for adding and deleting devices from its network. Furthermore, a UHG is responsible for transferring collected data to the cloud through network layer. In our testbed, UHG is built with Raspberry pi computer (Model-B Rev 1) that can communicate with sensors, actuators, smart plugs, and smart meters through different communications protocols including Zigbee, Z-wave and Bluetooth realising interoperability as shown in Fig.~\ref{fig:msg}. Please note that the gateway, which is a IP based system and runs in HTTP and XMPP protocols, serves as a translator and can communicate with other non-IP based devices (not only limited to BLE, Z-wave and Zigbee) in the system. Smart plugs that act as power outlet to different appliances communicate to UHG though Z-wave. Energy usage is pushed by UHG at regular interval and control command is pushed to the plug when necessary. Sensors like temperature, humidity, motion, light, noise update contextual and environmental data, and actuators such as IR Blasters and LED Potentiometers enable control systems. They communicate with UHG though Zigbee communication module and are powered by Arduino Fio micro controller. While traditional Zigbee protocol allows the Zigbee end device to push reading to the Zigbee gateway (i.e., UHG in our case), we have implemented a Z-wave like protocol to pull/push reading/control command to Zigbee end device to better manage the devices across different protocols and better control on delay of control command. Similarly, we perform for Bluetooth based sensors.

UHG establishes a two-way communication channel where it listens to control commands from cloud on one thread and sends periodic data collected by the smart devices on the other. When control information is received by UHG, it is pushed to end devices through NID by their respective communications protocols and enables the device with desired setting (e.g., set temperature of AC to $24^o$ C). Furthermore, while receiving asynchronous messages from Openfire server, we implement Mutex to prioritise asynchronous messages over periodic data polling, and thus prevent the smart devices from being confused or overwhelmed with requests.

The use of UHG per household eliminates the cloud server and Openfire to manage and control a large number of smart devices directly, which improves scalability of the system. Local storage and display can also be added to UHG such that consumers can retrieve historical energy usage directly from UHG instead from cloud. This reduces the burden of cloud and improves the scalability significantly.

\subsection{Consumer Side with Mobile Apps}
\subsubsection{GCM for real-time notifications}
We have used GCM to realise push on mobile side. GCM facilitates implementation of collaborative real-time applications for Android. GCM server manages real-time push notifications to the consumers even when they are offline (i.e., the apps is not activated on smart phone). When a consumer installs the app, they are registered with GCM and assigned a unique DID. This DID is recorded at our cloud server. When there is any update either on energy consumption or emergency, it will be pushed to the app for a specified user. For instance, when energy usage reaches a pre-set threshold, user gets notified.

For notifying consumers about dynamic pricing, we use HTTP multicast addressing supported by GCM and send messages to consumers.

\subsubsection{Mobile Application for user interface}
To utilize personal smart phone or tablet to control and monitor own smart devices in our IoT testbed, an Android mobile application was developed and installed on all consumers' smart phones. To display real-time information (e.g., Dynamic Pricing) and for user to send control command (e.g., switch on/off plug) to the UHG, native push to smart phone devices through GCM was implemented.

\section{Case Study, Results and Discussion}\label{sec:section6}
\subsection{Case Study}
Now, we demonstrate several case studies on how our platform is applicable to perform certain applications related to energy management.
\subsubsection{Demand Response Management}
\begin{figure*}[t]
\centering
\includegraphics[width=\textwidth]{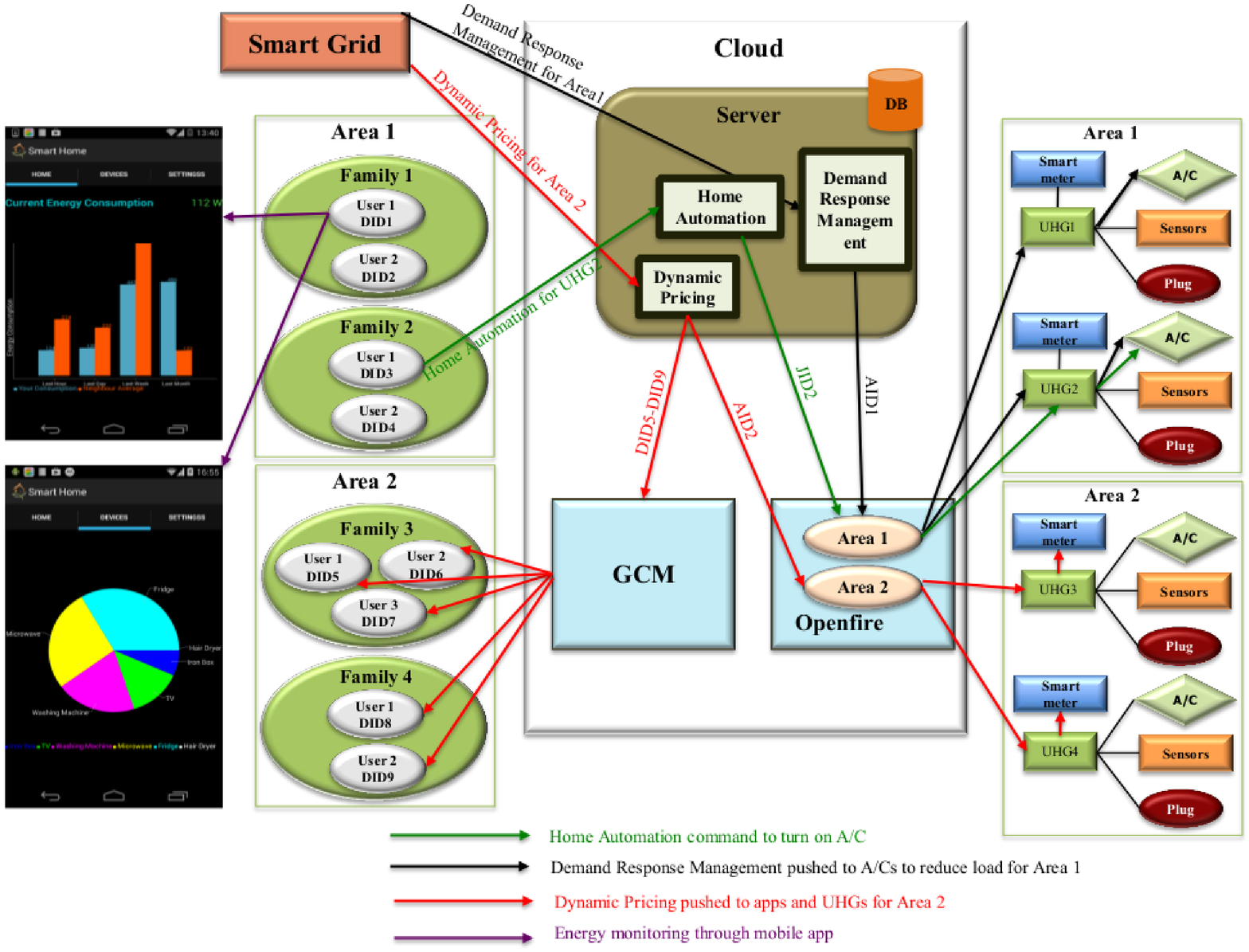}
\caption{Case studies for demand response management, dynamic pricing,
home automation, and energy monitoring. Here, the screen shots on the left hand side demonstrate how a user can view in real-time his total energy consumption as well as the consumed energy amount by each of the appliances in the home.}
\label{fig:dp}
\end{figure*}

Essential functionalities of DRM include load balancing during peak demand periods and locate and drive power to emergency hit areas. In the Fig.~\ref{fig:dp}, we illustrate our communication pattern for achieving DRM. In case of high energy demand, we cut down power consumption as our energy management system will notify grid service on cloud which in turn will publish command to all the gateways through Openfire and locate and route power around trouble spots to manage the situation. For instance, when smart grid detects a high demand and wants to reduce the excess demand, it will notify the DRM module on the cloud server, where the module may decide to have all AC units in Area 1 be adjusted with 2 degrees higher. Such control message will be forwarded to Openfire, and Openfire will notifies all UHGs in Area 1 through broadcast id AID1. Then, UHG will send the necessary control command to the AC unit.

\subsubsection{Dynamic Pricing}
To shape the demand and encourage energy awareness among consumers, our smart home system sends dynamic pricing information to consumers and smart meters. This greatly helps consumers to reduce energy usage during peak periods enabling demand shaping. Fig.~\ref{fig:dp} gives an overview of communication pattern for dynamic pricing for Area 2. Smart grid notifies our cloud and cloud in turn notifies XMPP servers via grid service to reach out to the consumers and smart meters, and hence encouraging consumers to run load consuming tasks during off-peak periods. Fig.~\ref{fig:dp} illustrates dynamic pricing flow in our energy management application where the dynamic pricing module in the cloud server broadcasts real-time pricing info to the affected area (e.g. AID2 in this case) once being notified by smart grid. At the same time, the cloud server also notifies users on their mobile through GCM (in this case DID5 - DID9 for the users in AID2).

\subsubsection{Energy Monitoring}
For energy monitoring, we focus on household appliances and monitor their usage to better understand the demand characteristics of the household and of the grid as a whole before making smart decisions on load scheduling and pricing. Fig.~\ref{fig:dp}  demonstrates our mobile application for energy monitoring, where User1 from Family 1 can monitor his energy consumption using mobile application that gives graphical data to the user.

\subsubsection{Home Automation}
Multiple consumers in a family can control the set of appliances through the home automation services. For instance, a user, while at his office, wants to switch AC in his home to a preferred temperature before his arrival. For this purpose, he (User 1 of Family 2 ) will first send a control command with desired AC setting via Android application. Upon receiving this command, home automation service on cloud will process it and push the message along with JID of the corresponding UHG to Openfire. As soon as Openfire receives this information, it will push it towards the corresponding UHG. Then, the UHG will process the message and set control setting of the respective AC according to the command. Note that the similar automation technique is applicable when a user control other smart devices in the house. Fig.~\ref{fig:dp}, illustrates home automation, where User 1 from Family 1 sends a command to the Home Automation module on the cloud server to control his AC system remotely. Server in turn will send the JID (JID2) to Openfire to control the AC system, which notifies the UHG and control it. We stress that the case study on home automation can be conducted in terms of other metrics such as energy savings. However, demonstration of such metrics requires a large discussion on other related topics such as algorithm for peak shaving~\cite{Wen-TaiAccess:2015}, real-time control algorithm and user study and incentive design~\cite{Naveed-Energies:2013,NaveedAccess:2015}, which are beyond the scope of this paper and hence are not discussed here.
\subsubsection{Home Security}
In case of Home Security, user informs our cloud when they plan to go on a holiday. If intrusion is detected during that period by data management service, it will activate home security service which in turn will send a warning message to all consumers in the family and to UHG to start camera to record environment at home. Users can take actions henceforth.

\subsection{Results and Discussion}
\begin{figure*}[t]
\centering
\includegraphics[width=\textwidth]{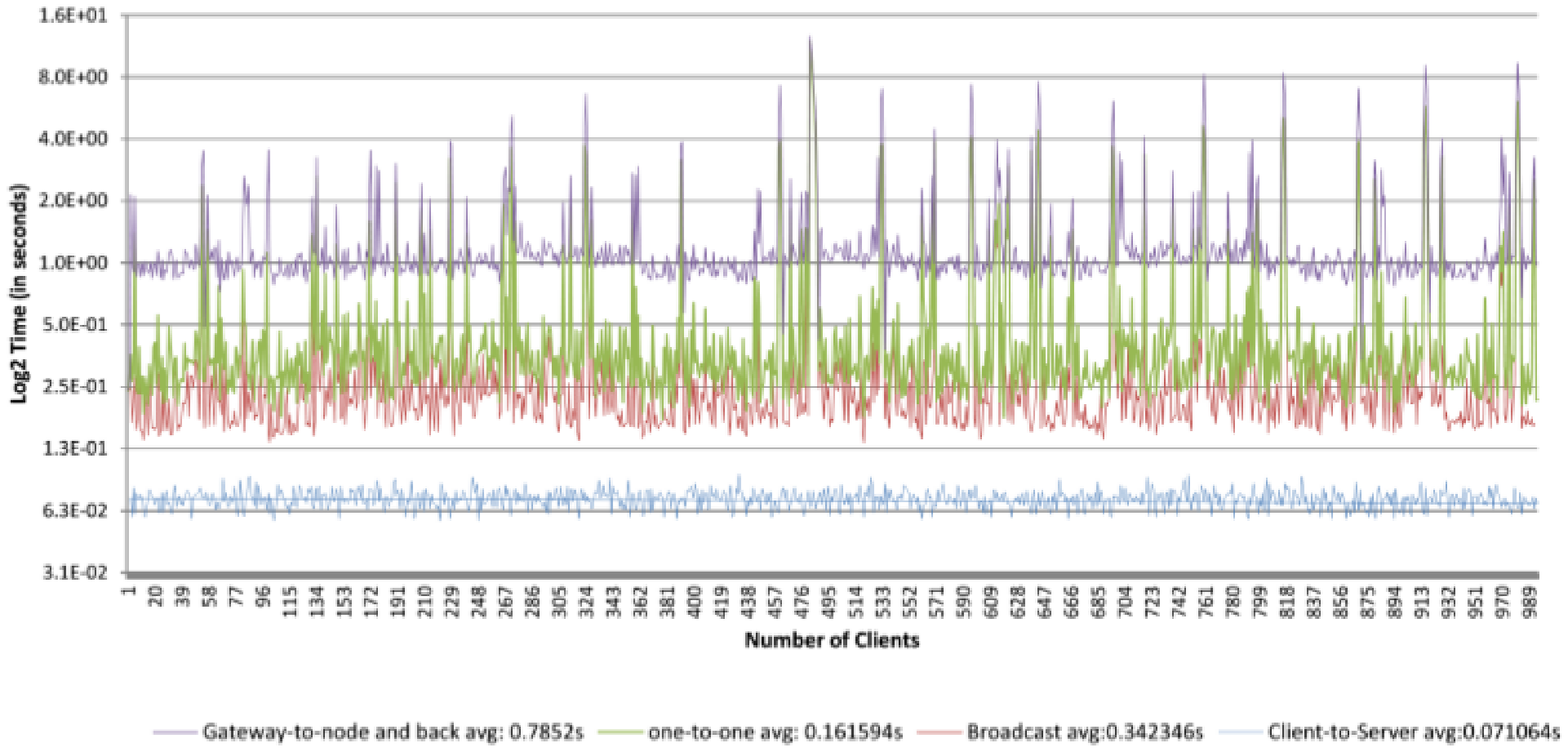}
\caption{Delay Test Results for 1000 clients}
\label{fig:res}
\end{figure*}

We performed delay and stress test to measure the time taken by each component in our system to deliver and process messages. We performed the test with 1000 simultaneous client requests and recorded the delay in terms of seconds. In our experimental set up, as illustrated in Fig.~\ref{fig:dp}, we are running the client (mobile app), cloud server and Openfire on one Local Area Network (LAN) and the UHG on another LAN, with all the nodes connected to gateway through Z-wave. Our aim is to get a rough idea on the relative timing requirements for each portion. Results can be easily extended to wide area network by adding additional delay.

In Fig.~\ref{fig:res}, we can see that latency is low for sending a message from client (mobile app) to cloud server, which averages to $0.07106363$s. However, average time required to control a node from gateway and get acknowledgement is $0.7852$s. A higher latency is due to Z-wave communication module and the limited processing speed of Raspberry Pi gateway. Further, broadcast message (through AID) and one-to-one message (through JID) from openfire to UHG takes $0.34234637$s and $0.1615937$s respectively. The delay is considerably small and is in terms of milliseconds (thus, less than $8$ seconds). We have shown that our system has lower latencies even when there are $1000$ simultaneous requests\footnote{We assume one request at a time to be processed by the server and to be received by the UHG as the processing speed is highly dependent on the processor power and other background load. Since we are interested in delay (especially each contribution of the delay) in our experiment, we assume just one request at time to remove other interruptions.}. However, latencies might be higher if the system needs to be connected in public networks.

\begin{figure}[t]
\centering
\includegraphics[width=\columnwidth]{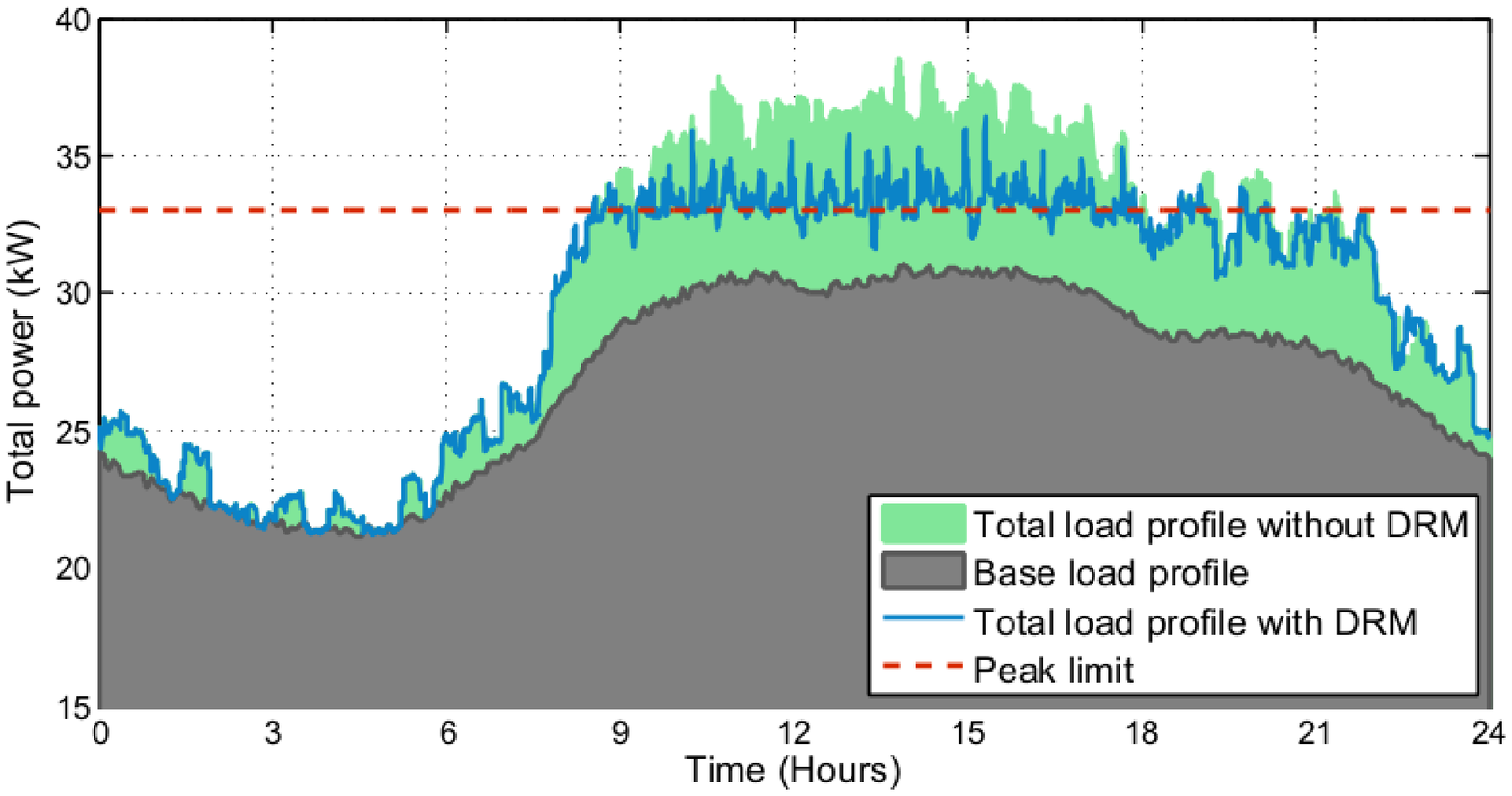}
\caption{Demonstration of peak load shaving through controlling loads of the homes using the proposed IoT architecture.}
\label{fig:fig2DRM}
\end{figure}

Now, to demonstrate the usefulness of proposed IoT architecture for energy management in smart grid we set up an experiment with 10 homes in a small residential community where each home is equipped with a UHG.  Lighting units of each home are considered as the controllable loads and assumed to have on and off control. The base load is assumed to be a random curve based on the reports of National Electricity Market of Singapore (NEMS). The daily usage pattern of the controllable loads are generated with their probability of turning on. Now, once the grid system detects a demand exceeding the peak level it informs the DRM unit to reduce the consumption loads with a view to keep the total demand lower than the maximum available limit. Then, the DRM unit decides on which lights should be turned off by solving an optimization problem. The detail of the optimization problem can be found in \cite{Wen-TaiAccess:2015}. For this experiment, the maximum allowable peak demand is assumed to be $33$ kW and DRM unit is notified to control the loads whenever the total consumption is greater than $33.5$ kW.

In Fig.~\ref{fig:fig2DRM}, we show the use of proposed IoT system in controlling the light bulbs at each house for reducing the peak demand in smart grid through DRM\footnote{Please note that shifting of appliance from peak hours to off-peak hours could be another potential way to perform DRM. Examples of such scheme can be found in \cite{Wen-TaiAccess:2015},  \cite{NaveedAccess:2015} and \cite{Naveed:2015}. However, this is beyond the scope of this paper.}. In Fig.~\ref{fig:fig2DRM}, $24$ hours time duration is divided into $8640$ time slots, i.e., the smart grid detects the total demand in every 10 seconds. The green zone and blue zone of the figure denote the load profile with and without demand management respectively whereby grey zone indicate the base load. The red line shows the maximum allowable peak demand threshold ($33$ kW). As demonstrated in Fig.~\ref{fig:fig2DRM}, when the total demand is under peak limit, smart grid does not need to do any controlling and hence no difference is observed between green zone and the blue line. However, once the total demand exceeds the peak limit, DRM unit controls the lights and turns some of them off to reduce demand load promptly as indicated by the blue line during $9-18$ o'clock. Thus, the result in Fig.~\ref{fig:fig2DRM} clearly shows that our IoT platform is effective in preforming energy management applications. However, we stress that in some instants the total load exceeds the peak limits during DRM. This is due to the fact that the smart power meter samples the total load at each $10$ seconds interval. If demand increases rapidly within these $10$ seconds, a delay occurs when the server receives the total load data to perform DRM. Therefore, the total load exceeds the peak limit sometimes. Furthermore, while controlling some appliances for managing the excess demand, there could be some new loads that are switched on by the users that also contribute to the overall demand. Hence, total load exceeds the threshold. However, the loads fall back below the peak limit promptly after performing DRM.

\section{Conclusion}\label{sec:section7}
In this paper we have given an overview of IoT elements along with architectural layers, compare IoT protocols like RESTful HTTP, CoAP, MQTT, and XMPP and IoT features that are customised to smart grid applications, such as security, provisioning, interoperability, latency and scalability. Following the design principles and IoT software protocols and features, we have developed a testbed where each housing unit is equipped with sensors, actuators, smart plugs, smart meters and UHGs. We have also highlighted the implementation of our applications in DRM, energy management, home automation, dynamic pricing and home security, which are critical factors in addressing efficient energy management and enabling a smarter lifestyle for consumers.


\begin{thebibliography}{10}
\providecommand{\url}[1]{#1}
\csname url@samestyle\endcsname
\providecommand{\newblock}{\relax}
\providecommand{\bibinfo}[2]{#2}
\providecommand{\BIBentrySTDinterwordspacing}{\spaceskip=0pt\relax}
\providecommand{\BIBentryALTinterwordstretchfactor}{4}
\providecommand{\BIBentryALTinterwordspacing}{\spaceskip=\fontdimen2\font plus
\BIBentryALTinterwordstretchfactor\fontdimen3\font minus
  \fontdimen4\font\relax}
\providecommand{\BIBforeignlanguage}[2]{{%
\expandafter\ifx\csname l@#1\endcsname\relax
\typeout{** WARNING: IEEEtran.bst: No hyphenation pattern has been}%
\typeout{** loaded for the language `#1'. Using the pattern for}%
\typeout{** the default language instead.}%
\else
\language=\csname l@#1\endcsname
\fi
#2}}
\providecommand{\BIBdecl}{\relax}
\BIBdecl

\bibitem{Ruilong-TII:2015}
R.~Deng, Z.~Yang, M.-Y. Chow, and J.~Chen, ``A survey on demand response in smart grids: {M}athematical models and approaches,'' \emph{IEEE Trans. Ind. Informat.}, vol.~11, no.~3, pp. 570--582, June 2015.

\bibitem{Tushar-TIE:2014}
W.~Tushar, B.~Chai, C.~Yuen, D.~B. Smith, K.~L. Wood, Z.~Yang, and H.~V. Poor, ``Three-party energy management with distributed energy resources in smart grid,'' \emph{IEEE Trans. Ind. Electron.}, vol.~62, no.~4, pp. 2487--2498,  Apr. 2015.

\bibitem{Liu:2014}
Y.~Liu, C.~Yuen, S.~Huang, N.~Ul~Hassan, X.~Wang, and S.~Xie, ``Peak-to-average ratio constrained demand-side management with consumer's preference in residential smart grid,'' \emph{IEEE J. Sel. Topics Signal Process.}, vol.~8, no.~6, pp. 1084--1097, Dec 2014.

\bibitem{Liu-TSG:2015}
Y.~Liu, C.~Yuen, R.~Yu, Y.~Zhang, and S.~Xie, ``Queuing-based energy consumption management for heterogeneous residential demands in smart grid,''  \emph{IEEE Trans. Smart Grid}, vol. Pre-print, pp. 1--10, 2015, (DOI: 10.1109/TSG.2015.2432571).

\bibitem{IoTServey:2010}
L.~Atzori, A.~Iera, and G.~Morabito, ``{The internet-of-things: A survey},'' \emph{Computer Networks}, vol.~54, no.~15, pp. 2787--2805, Oct. 2010.

\bibitem{IoTServeystandard:2013}
M.~R. Palattella, N.~Accettura, X.~Vilajosana, T.~Watteyne, L.~A. Grieco, G.~Boggia, and M.~Dohler, ``{Standard protocol stack for the internet of (important) things},'' \emph{IEEE Commun. Surveys Tuts.}, vol.~15, no.~3, pp. 1389--1406, Third Quarter 2013.

\bibitem{RongYu:2015}
R.~Yu, J.~Ding, W.~Zhong, Y.~Zhang, S.~Gjessing, A.~Vinel, and M.~Jonsson, ``Price-based energy control for {V2G} networks in the industrial smart grid,'' in \emph{Proc. International Conference on Industrial Networks and Intelligent Systems (INISCom)}, Tokyo, Japan, Mar. 2015, pp. 107--112.

\bibitem{TusharWCM:2016}
W.~Tushar, C.~Yuen, B.~Chai, S.~Huang, K.~L. Wood, S.~G. Kerk, and Z.~Yang, ``Smart grid testbed for demand focused energy management in end user environments,'' \emph{IEEE Wireless Commun.}, vol. Pre-print, pp. 1--10, 2016, (available: http://arxiv.org/abs/1603.06756).

\bibitem{Xu2015}
L.~Xu, C.~Jiang, Y.~Chen, Y.~Ren, and K.~J.~R. Liu, ``Privacy or utility in data collection? {A} contract theoretic approach,'' \emph{IEEE J. Sel. Topics Signal Process.}, vol.~9, no.~7, pp. 1256--1269, Oct. 2015.

\bibitem{Keoh2014}
S.~L. Keoh, S.~S. Kumar, and H.~Tschofenig, ``Securing the internet of things: {A} standardization perspective,'' \emph{IEEE Internet Things J.}, vol.~1, no.~3, pp. 265--275, June 2014.

\bibitem{Guo2015}
L.~Guo, J.~Wu, Z.~Xia, and J.~Li, ``Proposed security mechanism for {XMPP}-based communications of {ISO/IEC/IEEE} 21451 sensor networks,'' \emph{IEEE Sensors J.}, vol.~15, no.~5, pp. 2577--2586, May 2015.

\bibitem{Wen-TaiAccess:2015}
W.-T. Li, C.~Yuen, N.~U. Hassan, W.~Tushar, and C.-K. Wen, ``Demand response management for residential smart grid: {F}rom theory to practice,'' \emph{IEEE Access (Special issue on Smart Grids: A Hub of Interdisciplinary Research)}, vol.~3, pp. 2431--2440, Nov. 2015.

\bibitem{Naveed-Energies:2013} 
N.~U. Hassan, M.~A. Pasha, C.~Yuen, S.~Huang, and X.~Wang, ``Impact of scheduling flexibility on demand profile flatness and user inconvenience in residential smart grid system,'' \emph{Energies}, vol.~6, no.~12, pp. 6608--6635, Dec 2013.

\bibitem{NaveedAccess:2015}
A.~Naeem, A.~Shabbir, N.~U. Hassan, C.~Yuen, A.~Ahmed, and W.~Tushar, ``Understanding customer behavior in multi-tier demand response management program,'' \emph{IEEE Access (Special issue on Smart Grids: A Hub of Interdisciplinary Research)}, vol.~3, pp. 2613--2625, Nov. 2015.

\bibitem{Naveed:2015}
N.~U. Hassan, Y.~Khalid, C.~Yuen, and W.~Tushar, ``Customer engagement plans for peak load reduction in residential smart grids,'' \emph{IEEE Trans. Smart Grid}, vol.~6, no.~6, pp. 3029--3041, Nov. 2015.

\end{thebibliography}
\end{document}